\documentclass[preprint,aps,nofootinbib]{revtex4}
\usepackage{graphicx}
\usepackage{amsmath}
\usepackage{amsfonts}
\usepackage{amssymb}
\usepackage{color}%
\providecommand{\U}[1]{\protect\rule{.1in}{.1in}}

\newcommand{\f}{\begin{equation}}
\newcommand{\ff}{\end{equation}}
\newcommand{\fa}{\begin{eqnarray}}
\newcommand{\ffa}{\end{eqnarray}}

\begin{document}
\title{The St\"{u}ckelberg Holographic Superconductors in Constant External Magnetic Field}
\author{Jian-Pin Wu}
\email{jianpinwu@yahoo.com.cn} \affiliation{Department of Physics, Beijing Normal University, Beijing 100875, China}

\begin{abstract}

We investigate the St\"{u}ckelberg holographic superconductor in present of the constant external magnetic field. We observe that a critical value of magnetic
field exists as the cases in usual holographic superconductor. Furthermore, we find that the external magnetic field strongly influence the phase transition of this model and have a jump in the condensate at the critical temperature even for $c_{4}=1$.

\end{abstract}
\maketitle
%===============================================================================
\section{Introduction}
%===============================================================================

The AdS/CFT correspondence \cite{MaldacenaADS,GubserADS,WittenADS,MaldacenaReviewADS} has become a powerful tool to explore strong coupled gauge theories, which maps a strongly coupled quantum field theory to a classical gravity description.
Therefore we can use the gravitational techniques to study field theories.
Recently, these techniques have also been applied in condensed matter physics (for reviews, see Refs.\cite{HorowitzReview,HartnollReview1,HartnollReview2,HerzogReview,McGreevyReview}).
Such applications origin from the discovery in Ref.\cite{OriginalDiscovery1,OriginalDiscovery2},
where Gubser pointed out that in an Abelian Higgs model coupled
to gravity in AdS space, Abelian symmetry of Higgs is spontaneously broken by the existence of black hole.
Base on this observation, a holographic superconductors is proposed by Hartnoll,
Herzog and Horowitz \cite{HHHPRLSC}. In such a simple model, they consider a charged scalar and the only Maxwell sector $A=A_{t}$ in a AdS black hole background.
This simple model capture some basic basic features of a superconductor
such as the existence of a critical temperature.
Furthermore, it has also been investigated extensively with a external magnetic field \cite{HHHJHEPSC,magnetic1,magnetic2,magnetic3,magnetic4,magnetic5,magnetic6,magnetic7,magnetic8}, where the spatially dependent Maxwell sectors have to be
considered.

Recently, Franco $et$ $al.$ proposed a simple generalization of the basic holographic superconductor model in
which the spontaneous breaking of a global $U(1)$ symmetry occurs via the St\"{u}ckelberg
mechanism\cite{stuckelberg1,stuckelberg2}. Here we will call it as the St\"{u}ckelberg holographic superconductor.
The St\"{u}ckelberg holographic superconductor can provide a description of a wider class of phase transitions.
One can tune the parameter to control some characteristic quantity in this system,
such as the strength of fluctuations, the size and strength of the coherence peak.
Furthermore, Q. Y. Pan and B. Wang study the St\"{u}ckelberg holographic superconductor in Einstein-Gauss-Bonnet gravity in Ref.\cite{GB}.
They find that the Gauss-Bonnet constant can also tune the order of the phase transition.
In this paper, we will study the behaviors of
the St\"{u}ckelberg holographic superconductor in the presence of a constant external magnetic field.
We find that the critical magnetic field exist as the cases in the usual holographic superconductor.
We will also show that how the parameter influence the critical magnetic field.

The outline of our paper is the following. In section II,
We firstly study the St\"{u}ckelberg holographic superconductor when a constant external magnetic field is added.
In section III, we obtain numerical solution of this model and explore the effects of the external magnetic field and other model
parameters on phase transition. The conclusion and discussion are given in section IV.

%===============================================================================
\section{The St\"{u}ckelberg Holographic Superconductors in Constant External Magnetic Field}
%===============================================================================

In this paper, we will consider the generalized St\"{u}ckelberg holographic superconductor\cite{stuckelberg2},
which include a $U(1)$ gauge field and the scalar field via a generalized St\"{u}ckelberg mechanics.
In this model, the action in an asymptotically $AdS_{4}$ background with a black hole is
\begin{eqnarray}
\label{stckelbergaction}
S=\int d^{4}\, \sqrt{-g}\,\Big\{-\frac{F_{\mu\nu}F^{\mu\nu}}{4}
-\frac{\partial_{\mu}\tilde{\Psi}\partial^{\nu}\tilde{\Psi}}{2}
-\frac{m^{2}}{2}\tilde{\Psi}^{2}
-\frac{1}{2}\mathcal{F}(\tilde{\Psi})(\partial_{\mu}p-A_{\mu})(\partial^{\mu}p-A^{\mu})\Big\}\ .
\end{eqnarray}
where $\mathcal{F}$ is a general function of $\tilde{\Psi}$
\begin{eqnarray}
\label{functionofF}
\mathcal{F}(\tilde{\Psi})=\tilde{\Psi}^{2}+c_{\alpha}\tilde{\Psi}^{\alpha}+c_{4}\tilde{\Psi}^{4}.
\end{eqnarray}
with $3 \leq \alpha \leq 4$. When $c_{\alpha} = c_{4} = 0$ it reduces to the model of \cite{HHHPRLSC}.
We can use the gauge freedom to fix $p=0$. In addition, we will take the ansatz $\tilde{\Psi}=\Psi(r)$, $A_{t}=\Phi(r)$.

When a external magnetic field is included, the solution of magnetically charged black hole in $AdS_{4}$ is given by \cite{solutionofMF}
\begin{eqnarray}
\label{metric}
ds^{2}=-f(r)dt^{2}+\frac{dr^{2}}{f(r)}+r^{2}(dx^{2}+dy^{2}).
\end{eqnarray}
\begin{eqnarray}
\label{functionoff}
f(r)=\frac{r^{2}}{L^{2}}-\frac{M}{r}+\frac{H^{2}}{r^{2}}.
\end{eqnarray}

Then the Hawking temperature of the RNAdS black hole can be expressed as
\begin{eqnarray}
\label{HawkingT}
T=\frac{f'(r_{+})}{4\pi}.
\end{eqnarray}
where $r_{+}$ is is the radius of the outer horizon of the RNAdS black hole, which is the most positive root of $f(r) = 0$.
Following \cite{stuckelberg2}, we shall consider the probe limit
in which the gauge field and scalar field do not back react on the geometry.
So, the equations of motion can be derived from the action (\ref{stckelbergaction})
\begin{eqnarray}
\label{equationofmotion1}
\Psi''+\left(\frac{f'}{f}+\frac{2}{r}\right)\Psi'
-\frac{m^{2}}{f}\Psi+\frac{\dot{\mathcal{F}}(\Psi)}{2f^{2}}\Phi^{2}=0.
\end{eqnarray}
\begin{eqnarray}
\label{equationofmotion2}
\Phi''+\frac{2}{r}\Phi'-\frac{\mathcal{F}(\Psi)}{f}\Phi=0.
\end{eqnarray}
where the prime denotes derivative with respect to $r$ but the dot on the $\mathcal{F}$ is derivative with respect to $\Psi$ and we have set $L=1$.
As in \cite{HHHPRLSC}, we will choose the mass $m^{2}=-2$.
In order to solve numerically the equations of motion (\ref{equationofmotion1}) and (\ref{equationofmotion2}),
the appropriate boundary conditions must be required.
%These two equations can be solved numerically by doing integration from the horizon out to the infinity.
Firstly, at the horizon, the regularity gives two conditions:
\begin{eqnarray}
\label{BconditionH1}
f'(r_{+})\Psi'(r_{+})=m^{2}\Psi(r_{+}).
\end{eqnarray}
\begin{eqnarray}
\label{BconditionH2}
\Phi(r_{+})=0.
\end{eqnarray}
Close to the AdS boundary $(r \rightarrow \infty)$, the solutions are found to be
\begin{eqnarray}
\label{BconditionI1}
\Psi=\frac{\Psi_{1}}{r}+\frac{\Psi_{2}}{r^{2}}.
\end{eqnarray}
\begin{eqnarray}
\label{BconditionI2}
\Phi=\mu-\frac{\rho}{r}.
\end{eqnarray}
where $\mu$ and $\rho$ are the chemical potential and charge density, respectively. The scalar
condensate is given by $\langle\mathcal{O}_{i}\rangle=\Psi_{i}, i=1,2$. Notice that for $\Psi$, both of these falloffs are
normalizable, so one can impose boundary condition that either $\Psi_{1}$ or $\Psi_{2}$ vanishes.
For simplicity, we will take $\Psi_{1}=0$. Thus, the scalar condensate is now described
by the operator $\langle\mathcal{O}_{2}\rangle=\Psi_{2}$ and we will discuss the condensate $\langle\mathcal{O}_{2}\rangle$ for fixed charge density (here we will set $\rho=-1$).

%===============================================================================
\section{The Numerical Analysis}
%===============================================================================

\begin{figure}[H]
\includegraphics[scale=0.6]{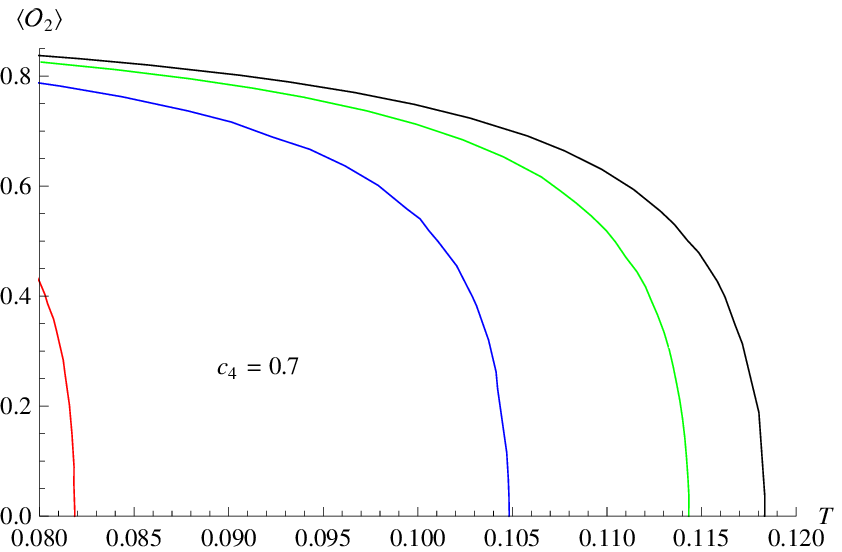}\hspace{0.2cm}%
\includegraphics[scale=0.6]{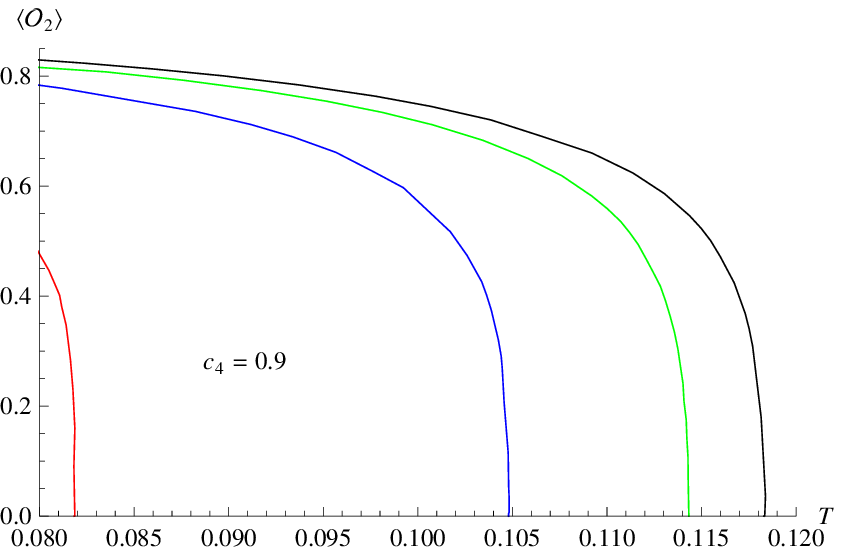}\\ \vspace{0.0cm}
\includegraphics[scale=0.6]{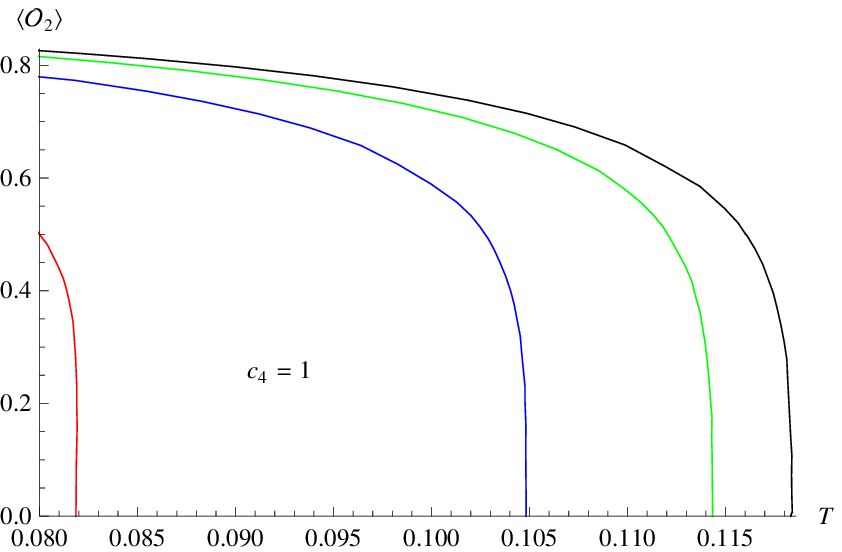}\hspace{0.2cm}%
\includegraphics[scale=0.6]{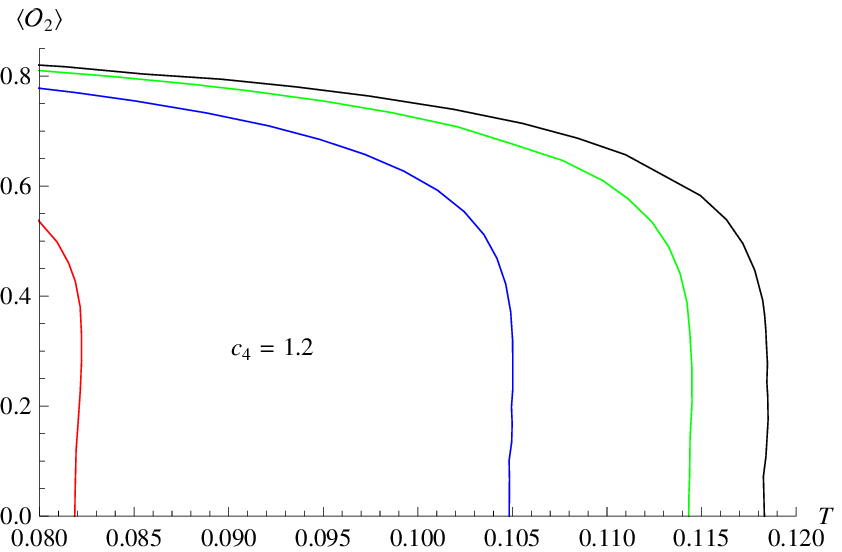}\hspace{0.2cm}%
\includegraphics[scale=0.6]{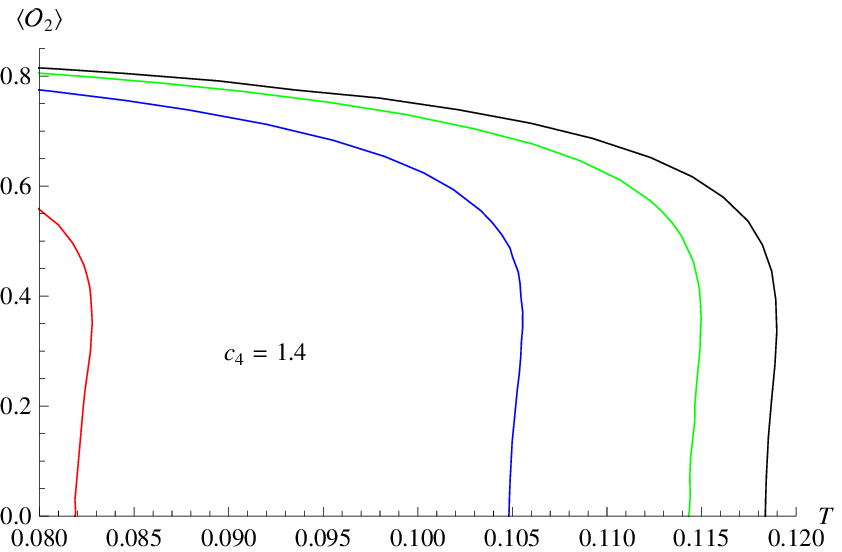}\\ \vspace{0.0cm}
\caption{\label{Condense1} The condensate
$<\mathcal{O}_{2}>$ as a function of temperature with selected values
$c_{4}$ ($c_{\alpha}=0$) for various values of $H^{2}$, $0.4$ (red), $0.25$
(blue), $0.1$ (green) and $0$ (black).}
\end{figure}
\begin{figure}[H]
\includegraphics[scale=1]{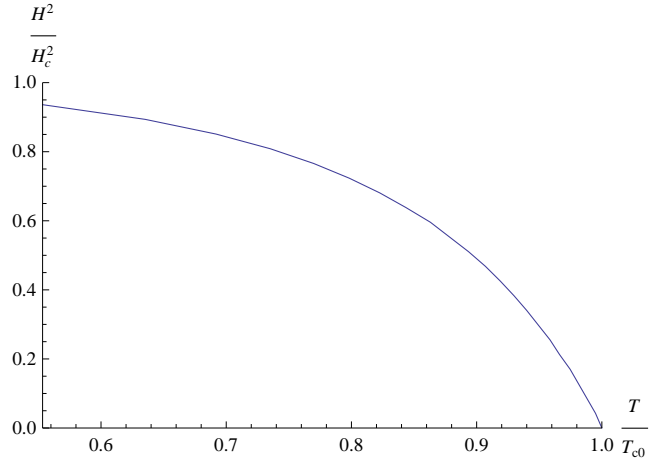}
\caption{\label{HT} The relations between the critical temperature $T_{c}$ on the
magnetic field $H^{2}$, where $T_{c0}$ and $H_{c}$ are the critical temperature in the absence of the magnetic field
$(H = 0)$ and the critical magnetic field at zero temperature, respectively.}
\end{figure}
\begin{figure}[H]
\includegraphics[scale=0.8]{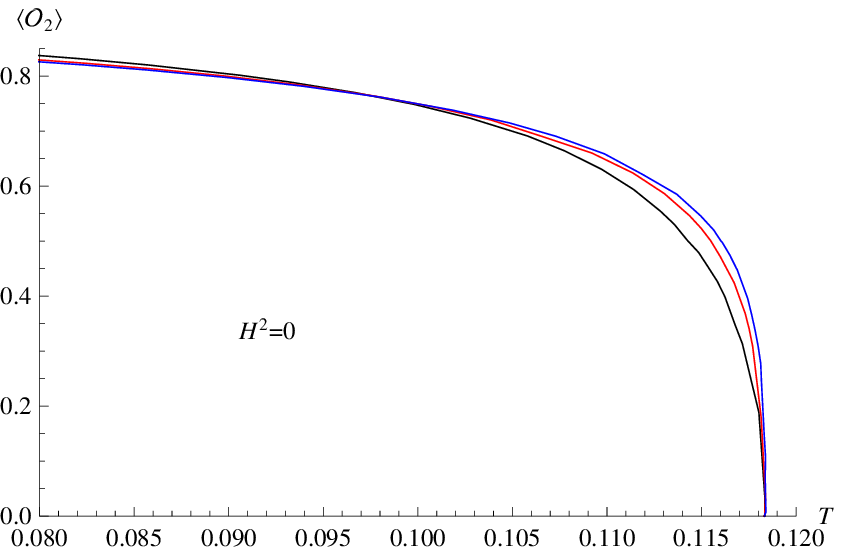}\hspace{0.2cm}%
\includegraphics[scale=0.8]{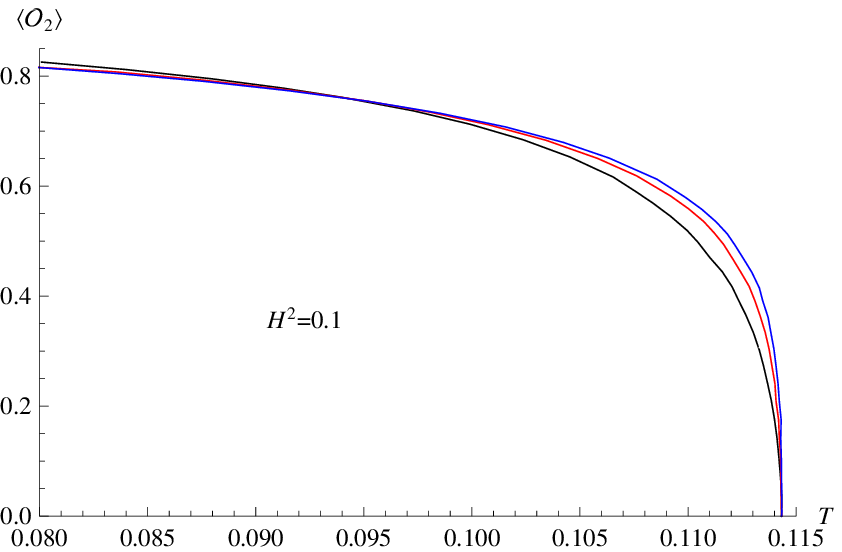}\\ \vspace{0.0cm}
\includegraphics[scale=0.8]{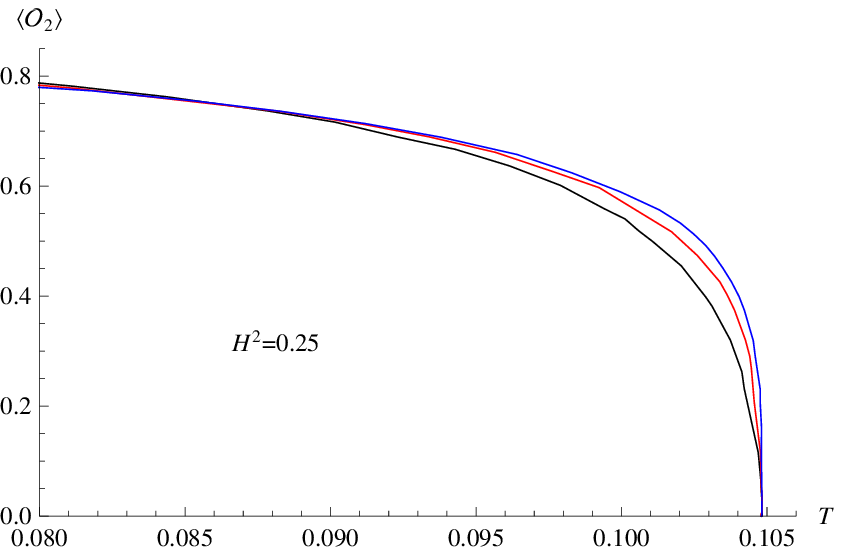}\hspace{0.2cm}%
\includegraphics[scale=0.8]{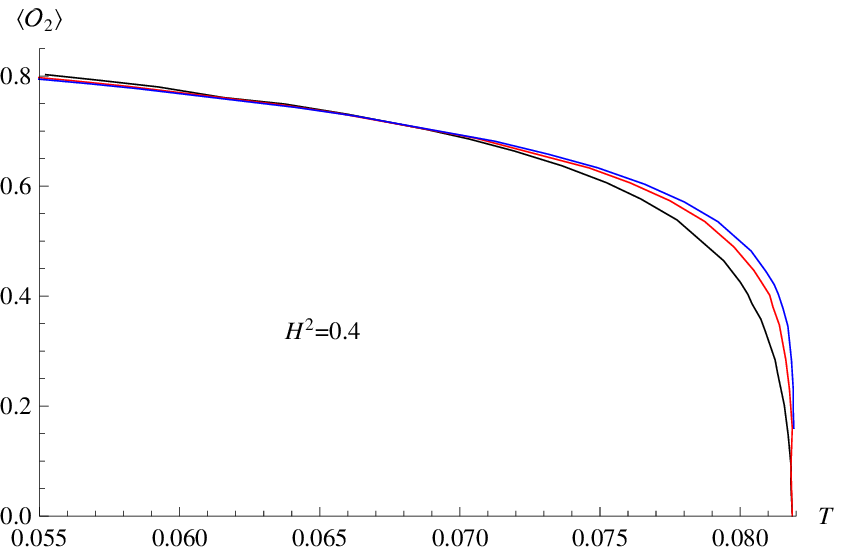}\\ \vspace{0.0cm}
\caption{\label{Cond2} $<\mathcal{O}_{2}>$ as a function of temperature with selected values of $H^{2}$ and various
$c_{4}$ ($c_{4}\leq1$), $0.7$ (black), $0.9$
(red) and $1$ (blue).}
\end{figure}
\begin{figure}[H]
\includegraphics[scale=0.8]{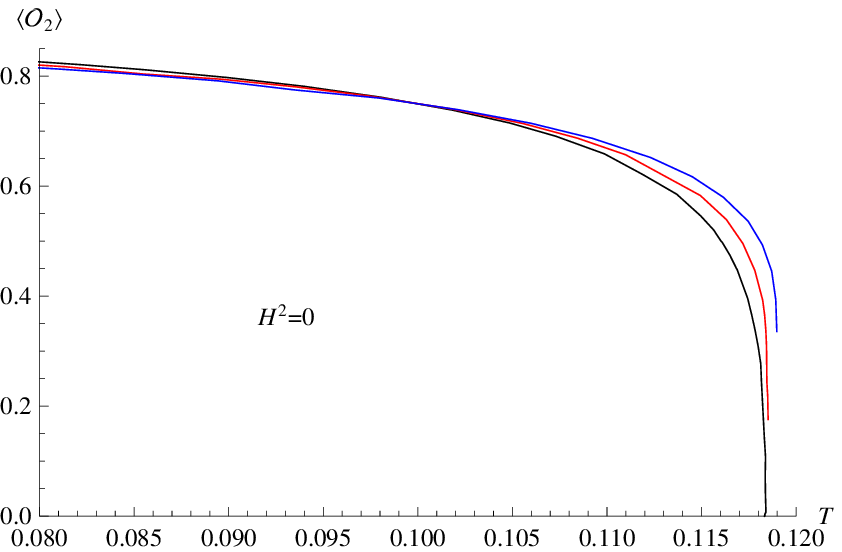}\hspace{0.2cm}%
\includegraphics[scale=0.8]{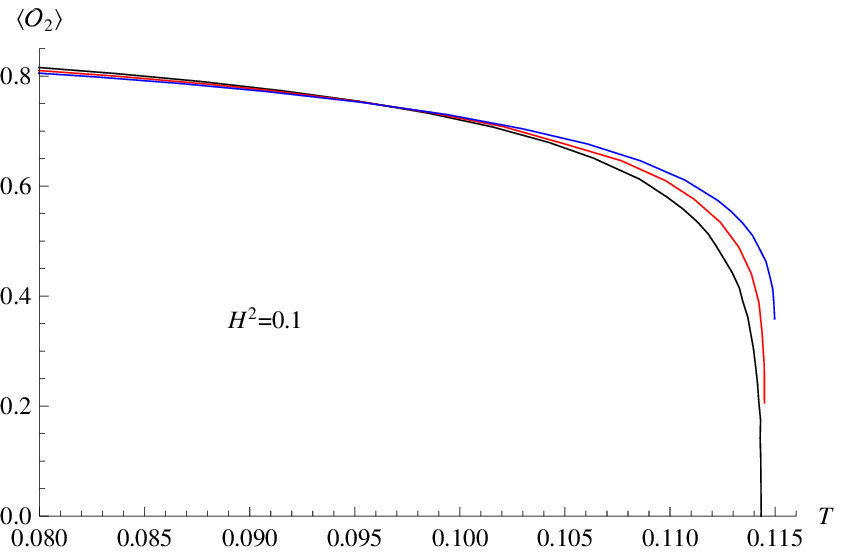}\\ \vspace{0.0cm}
\includegraphics[scale=0.8]{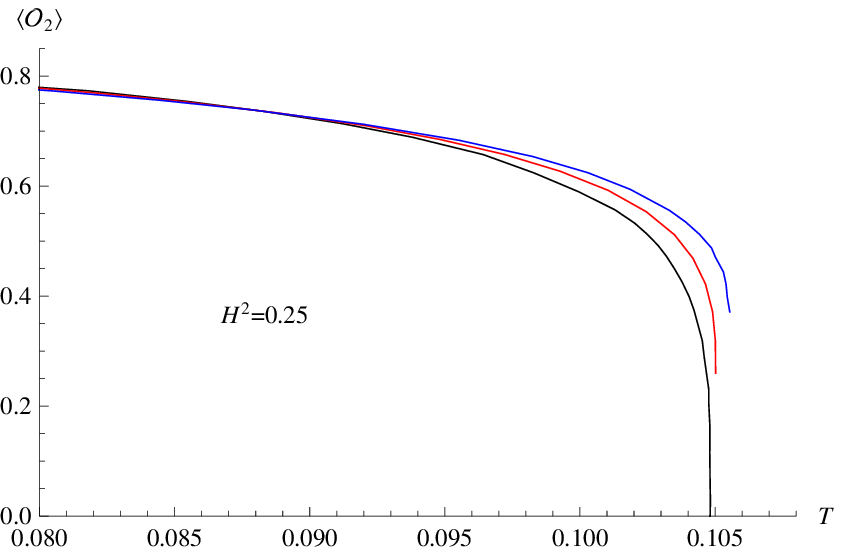}\hspace{0.2cm}%
\includegraphics[scale=0.8]{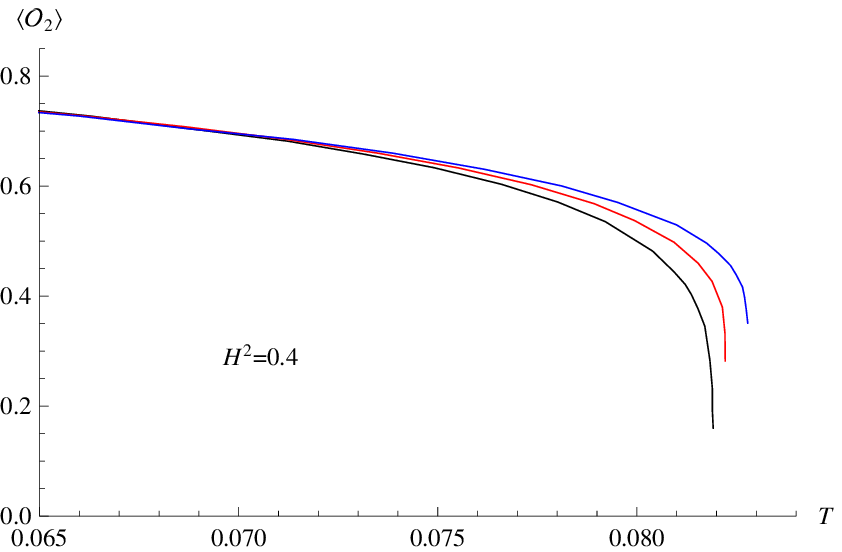}\\ \vspace{0.0cm}
\caption{\label{Cond3} $<\mathcal{O}_{2}>$ as a function of temperature with selected values of $H^{2}$ and various
$c_{4}$ ($c_{4}\geq1$), $1$ (black), $1.2$
(red) and $1.4$ (blue).}
\end{figure}

We will investigate the behaviors of the St\"{u}ckelberg holographic superconductor in the presence of a constant external magnetic field
and how the parameter $c_{\alpha}$ and $c_{4}$ influence the critical magnetic field.
The special attention on the case $\mathcal{F}(\Psi)=\Psi^{2}+c_{4}\Psi^{4}$ \cite{stuckelberg2,special1,special2,GB} have been payed, so we will firstly consider the case for $c_{\alpha}=0$ and pay more attention on the influence of $c_{4}$ on the phase transition. In addition, we also show the case for $c_{4}=0$ in order to explore the influence of $c_{\alpha}$ on the phase transition in the presence of the external magnetic field. We can straightforwardly
solve the equation of motion (\ref{equationofmotion1}) and (\ref{equationofmotion2}) numerically with the boundary conditions (\ref{BconditionH1}), (\ref{BconditionH2}), (\ref{BconditionI1}) and (\ref{BconditionI2}).

In FIG.\ref{Condense1} we plot the condensate $\langle\mathcal{O}_{2}\rangle$ for chosen values of $c_{4}$ and various values of the external magnetic field.
From the FIG.\ref{Condense1}, we can see that for different value of the parameter $c_{4}$, the critical temperature $T_{c}$ always decreases with the
increasing of the magnetic field $H$. This means that at a temperature slightly below $T_{0}$, the critical temperature at
zero magnetic field, there always exists a critical value of $H$, $H_{c}$, above which the material will be forced
back into the normal state from the superconductor state.
We also find that when the magnetic field $H$ is small, the decrease of
the critical temperature $T_{c}$ is very slow. However, when $H$ becomes lager
and larger, the critical temperature $T_{c}$ will decrease rapidly\footnote{In order for the black hole to be censored
by a horizon, the condition $27M^{4}-256H^{6}$ must be satisfied. $H^{2}_{max}=0.47247$ if we set $M=1$. For more discussions, we can refer to Refs. \cite{magnetic1,magnetic4}}.
Therefore, we can infer that when the magnetic field becomes large, it will strongly influence our holographic superconductor systems.
In FIG.\ref{HT}, we can also plot the relations between the critical temperature $T_{c}$ and the magnetic field $H$ for $c_{4}=1$.
We only plot the case of $T/T_{c0} \in [0.55,1]$. Beyond this interval, $H^{2}/H_{c}^{2}$ is close to $1$
(that is to say, the magnetic field is very large) and the numerical calculation will become difficult.
We have also examined the cases for $c_{4}=0.7,0.9$. We find that the curves of dependence of the critical temperature $T_{c}$ on the applied
magnetic field $H$ are same as the case for $c_{4}=1$.
However, beyond this interval, the cases may be very different because the magnetic field will strongly influence the phase transition and even have a jump in the condensate at the critical temperature for $c_{4}=1$ as we can see in the following.
In fact, when the parameter $c_{4} \leq 1$ and the magnetic field is small, varying the parameter $c_{4}$,
the critical temperature $T_{c}$ is invariable for the same external
magnetic field $H$. For illustration, we also plot the condensate as a function of temperature with chosen values of magnetic field for various values of $c_{4} \leq 1$ (FIG.\ref{Cond2}). However, when $c_{4} > 1$, the critical temperature $T_{c}$ changes with different values of $c_{4}$ even for the same applied magnetic field $H$ (FIG.\ref{Cond3}). Therefore, we can infer that when $c_{4} > 1$, the dependence of the critical temperature $T_{c}$ on the
magnetic field $H$ will be very different from the cases for $c_{4} \leq 1$. Finally, from FIG.\ref{Cond3} we have also observed that even for $c_{4}=1$, a jump in the condensate at the critical temperature also occurs while in the absence of the applied magnetic field, there is no jump for $c_{4}=1$ \cite{stuckelberg2}.
Therefore, the magnetic field can also tune the characteristic quantity of the St\"{u}ckelberg holographic superconductor as the other parameters of this model.

\begin{figure}[H]
\includegraphics[scale=0.6]{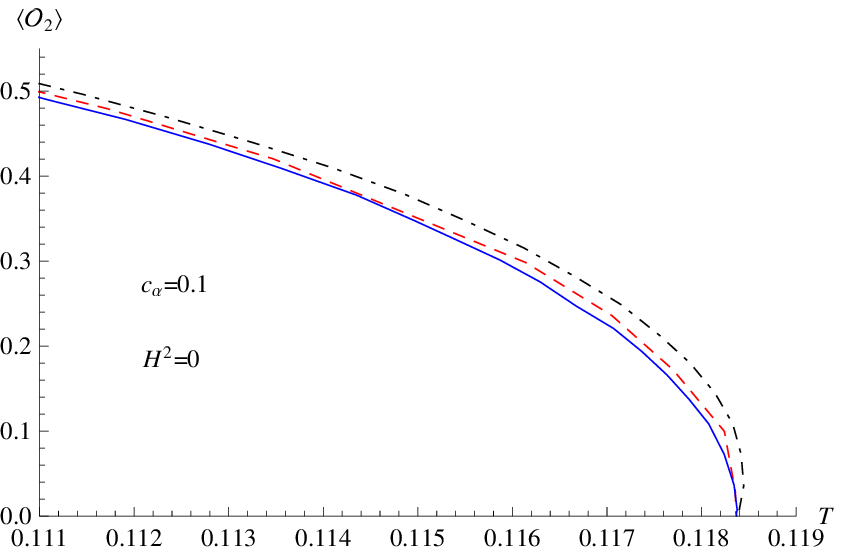}\hspace{0.2cm}%
\includegraphics[scale=0.6]{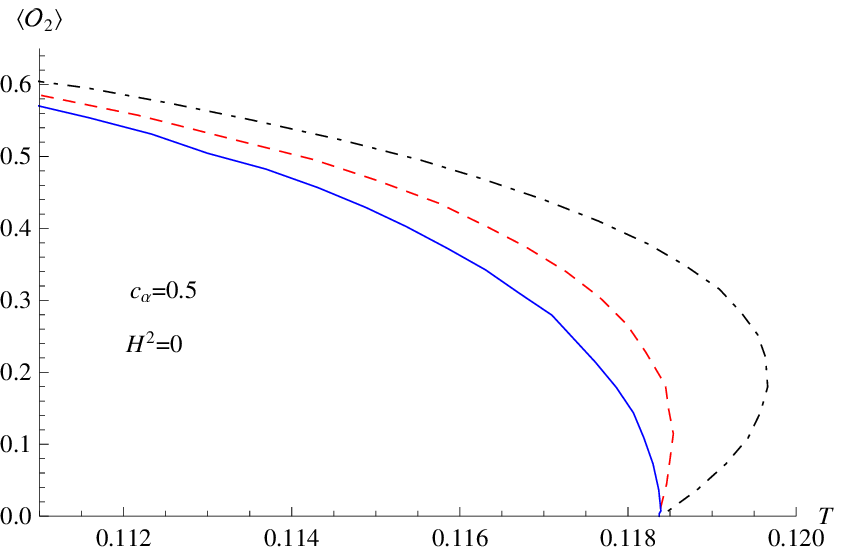}\hspace{0.2cm}%
\includegraphics[scale=0.6]{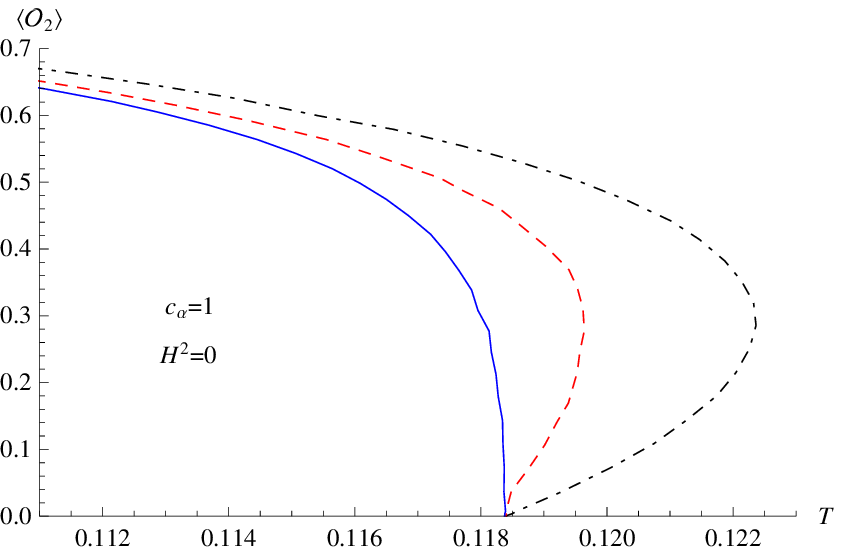}\\ \vspace{0.0cm}
\includegraphics[scale=0.6]{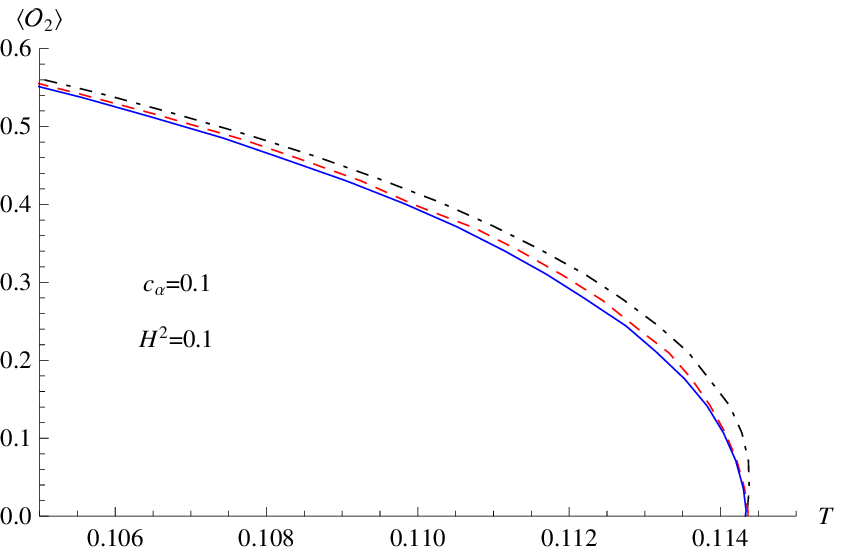}\hspace{0.2cm}%
\includegraphics[scale=0.6]{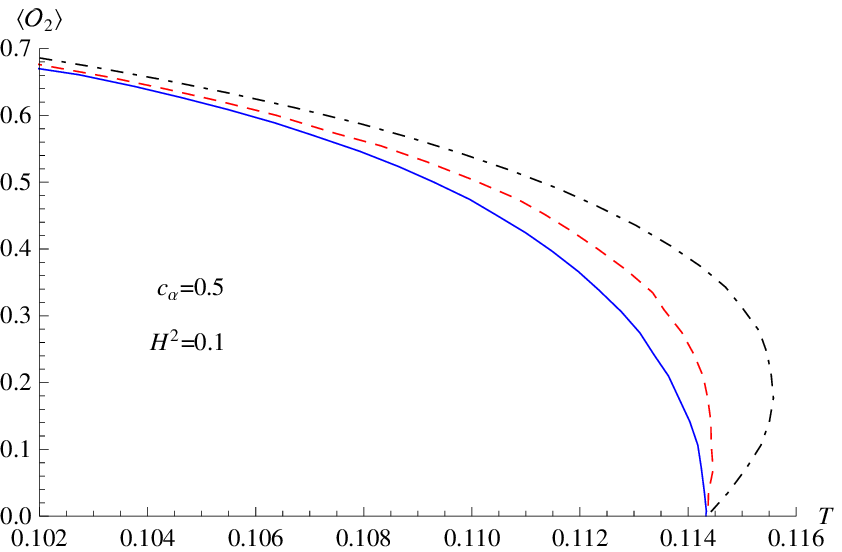}\hspace{0.2cm}%
\includegraphics[scale=0.6]{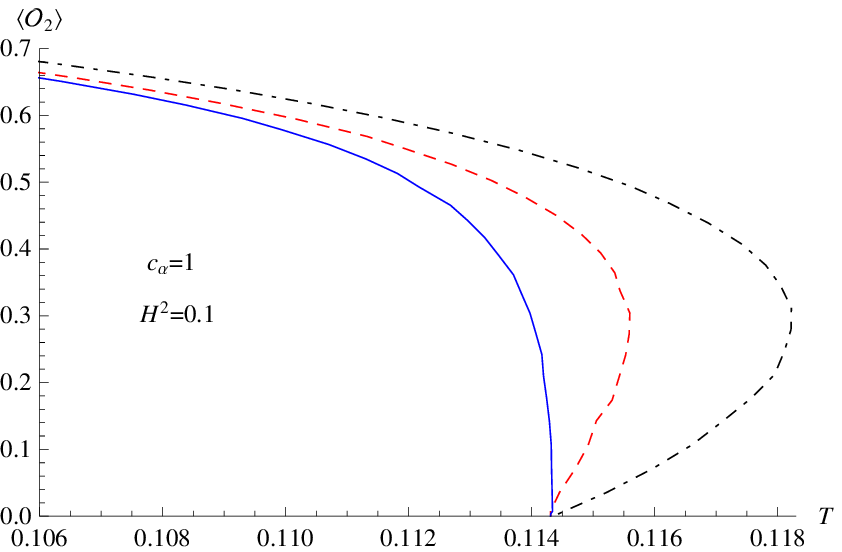}\\ \vspace{0.0cm}
\includegraphics[scale=0.6]{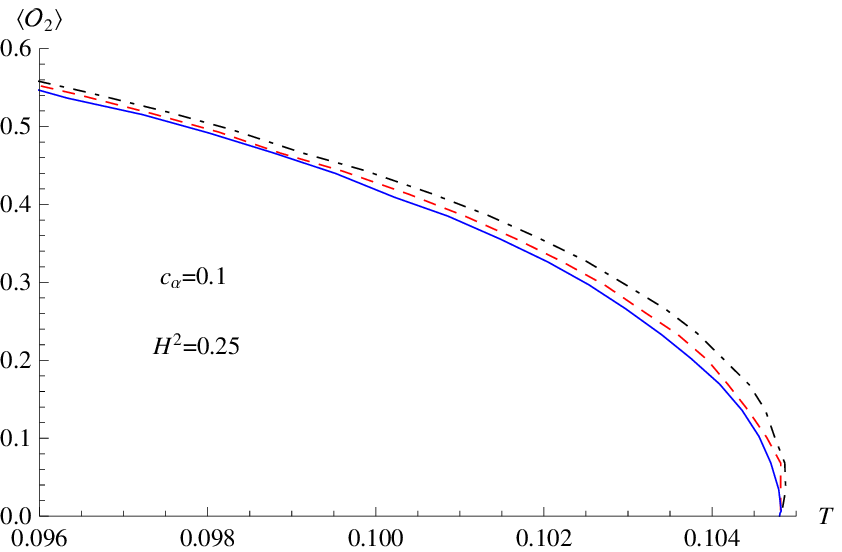}\hspace{0.2cm}%
\includegraphics[scale=0.6]{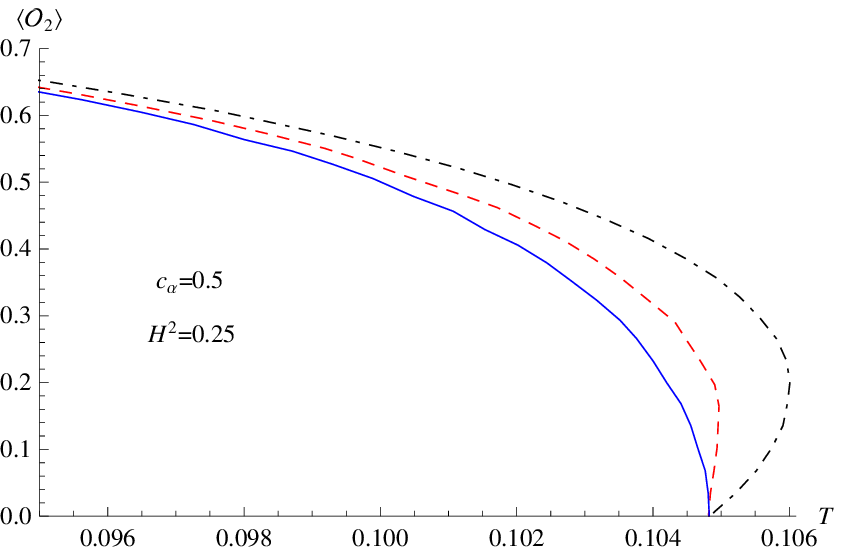}\hspace{0.2cm}%
\includegraphics[scale=0.6]{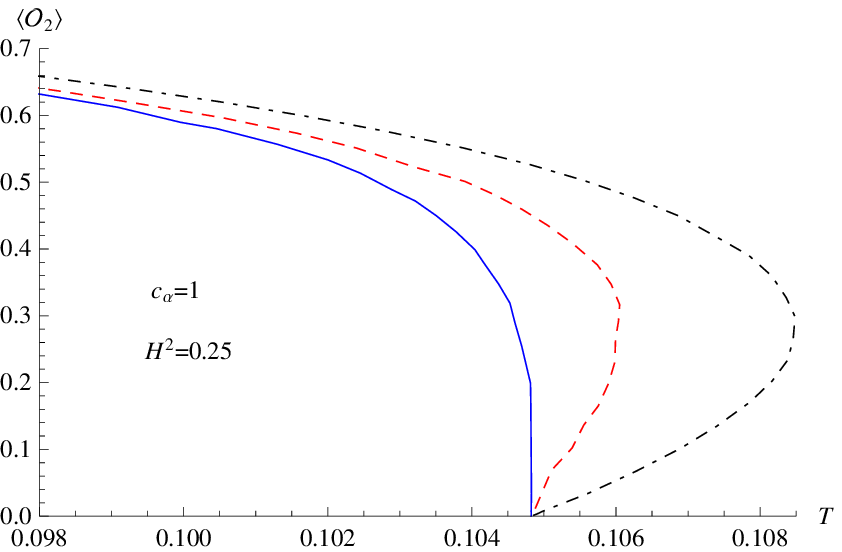}\\ \vspace{0.0cm}
\includegraphics[scale=0.6]{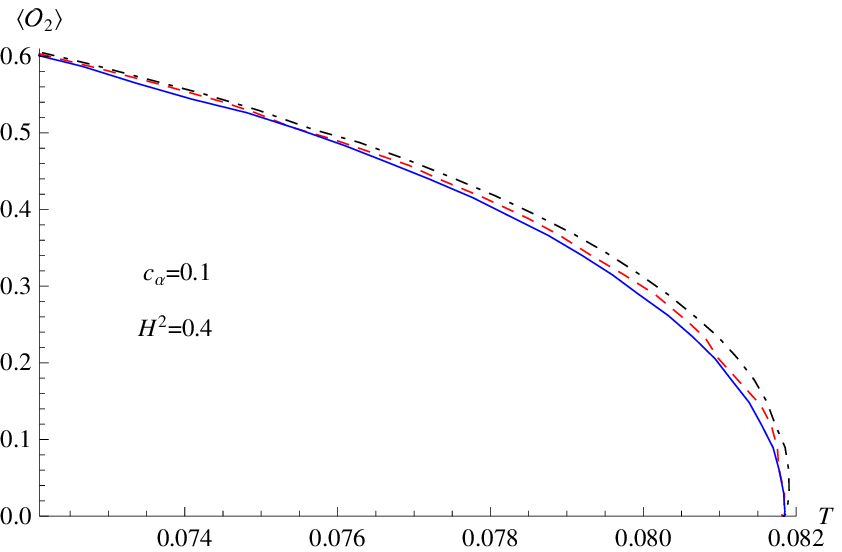}\hspace{0.2cm}%
\includegraphics[scale=0.6]{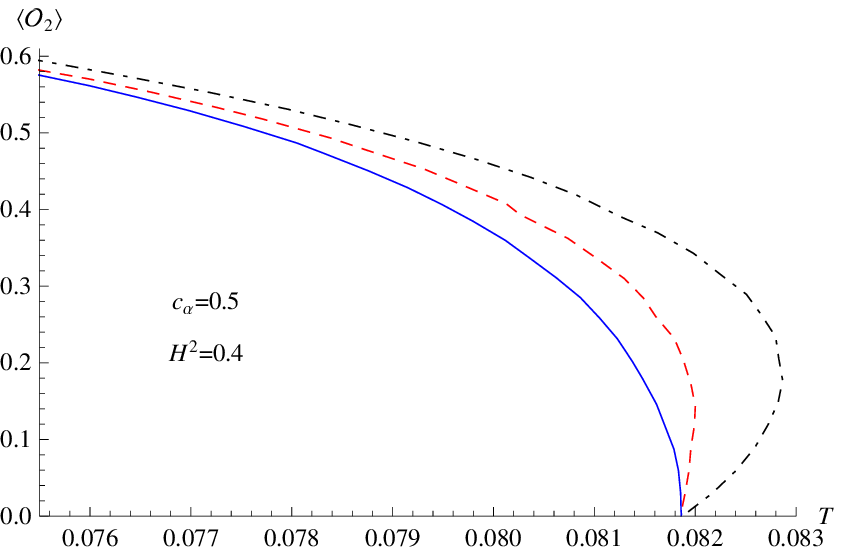}\hspace{0.2cm}%
\includegraphics[scale=0.6]{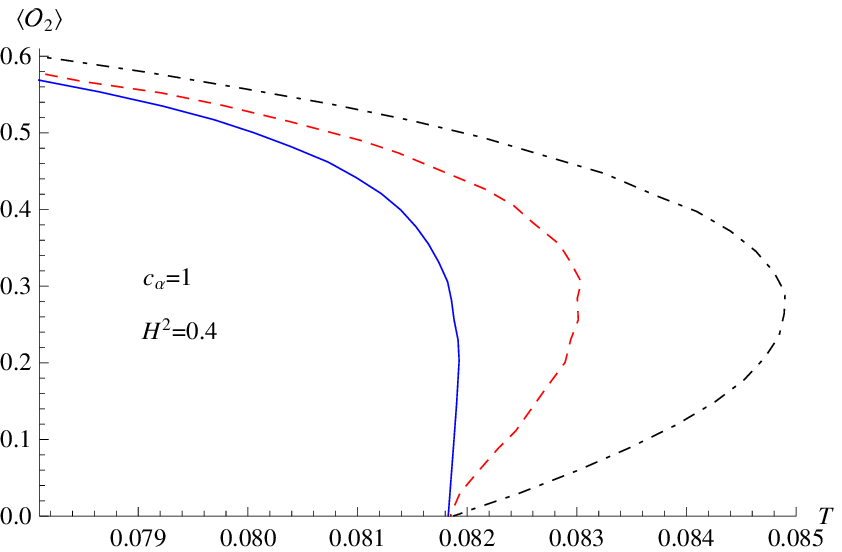}\\ \vspace{0.0cm}
\caption{\label{Condense5} The condensate
$<\mathcal{O}_{2}>$ as a function of temperature with selected values
$c_{\alpha}$ ($c_{4}=0$) and magnetic field for various values of $\alpha$, $3$ (black), $3.5$
(red) and $4$ (blue).}
\end{figure}

Subsequently, we shall explore the influence of $c_{\alpha}$ on the phase transition in the presence of the external magnetic field.
So we will set $c_{4}=0$. In FIG.\ref{Condense5}, the condensate $\mathcal{O}_{2}$ is plotted
with chosen values of $c_{\alpha}$ and magnetic field for various values of $\alpha$.
From FIG.\ref{Condense5}, we find that for the chosen values of $c_{\alpha}$ and magnetic field, the jump in the condensate at the critical temperature will appear as the parameter $\alpha$ increase. Also, the jump will be more obvious when the value of $c_{\alpha}$ becomes bigger.
In addition, we also note that the external magnetic field have slight influence on the appearance of the jump. The larger magnetic field, the jump is more obvious. It is consistent with the case for non-vanishing $c_{4}$ but vanishing $c_{\alpha}$.

%===============================================================================
\section{Conclusion and  Discussion}
%===============================================================================

We have studied the St\"{u}ckelberg holographic superconductor in the presence of the constant external magnetic field and we have obtained the numerical solution of this model. By the way of the numerical analysis, we find that the magnetic field will strongly influence the phase transition of this model and have a jump in the condensate at the critical temperature even for $c_{4}=1$ such that the magnetic field can also influence the characteristic quantity of the St\"{u}ckelberg holographic superconductor as the other parameters of this model. Therefore, we can infer that when the applied magnetic field is large, a richer physics in  will exist and deserve our furthermore investigations carefully.

In this paper, we only give a simple reports on the St\"{u}ckelberg holographic superconductor emerged in constant external magnetic field. Furthermore investigations are deserved, for example the Meissner effect or considering a less symmetric ansatz.

\section*{Acknowledgement}

This work is partly supported by NSFC(No.10975017).
The author also thanks Prof. Yi Ling and XiaoMei Kuang for exciting the investigation on holographic superconductors during his visit to Nanchang University.
He also thanks the correspondence on some special issues with Da-Zhu Ma.

\end{document}